\documentclass[prb,twocolumn,showpacs,superscriptaddress]{revtex4}
\usepackage[utf8]{inputenc}
\usepackage[T1]{fontenc}
\usepackage{graphicx}
\usepackage{dcolumn}
\usepackage{graphicx}
\usepackage{amsmath}
\usepackage{textcomp}
\usepackage{verbatim}
\usepackage[symbol]{footmisc}
\begin{document}
\title{Electroluminescence on-off ratio control of n-i-n GaAs/AlGaAs-based resonant tunneling structures}
\author{E. R. C. de Oliveira}
\affiliation{Departamento de Física, Universidade Federal de São Carlos, 13565-905, São Carlos, São Paulo, Brazil}
\affiliation{Technische Physik, Physikalisches Institut and Röntgen Center for Complex Material Systems (RCCM), Universität Würzburg, Am Hubland, D-97074 Würzburg, Germany}
\author{A. Pfenning}
\affiliation{Technische Physik, Physikalisches Institut and Röntgen Center for Complex Material Systems (RCCM), Universität Würzburg, Am Hubland, D-97074 Würzburg, Germany}
\author{E. D. Guarin}
\affiliation{Departamento de Física, Universidade Federal de São Carlos, 13565-905, São Carlos, São Paulo, Brazil}
\author{M. D. Teodoro}
\email[Corresponding author: ]{mdaldin@df.ufscar.br}
\affiliation{Departamento de Física, Universidade Federal de São Carlos, 13565-905, São Carlos, São Paulo, Brazil}
\author{E. C. dos Santos}
\affiliation{Departamento de Física, Universidade Federal de São Carlos, 13565-905, São Carlos, São Paulo, Brazil}
\author{V. Lopez-Richard}
\affiliation{Departamento de Física, Universidade Federal de São Carlos, 13565-905, São Carlos, São Paulo, Brazil}
\author{G. E. Marques}
\affiliation{Departamento de Física, Universidade Federal de São Carlos, 13565-905, São Carlos, São Paulo, Brazil}
\author{L. Worschech}
\affiliation{Technische Physik, Physikalisches Institut and Röntgen Center for Complex Material Systems (RCCM), Universität Würzburg, Am Hubland, D-97074 Würzburg, Germany}
\author{F. Hartmann}
\affiliation{Technische Physik, Physikalisches Institut and Röntgen Center for Complex Material Systems (RCCM), Universität Würzburg, Am Hubland, D-97074 Würzburg, Germany}
\author{S. Höfling}
\affiliation{Technische Physik, Physikalisches Institut and Röntgen Center for Complex Material Systems (RCCM), Universität Würzburg, Am Hubland, D-97074 Würzburg, Germany}
\affiliation{SUPA, School of Physics and Astronomy, University of St. Andrews, St. Andrews KY16 9SS, United Kingdom}


\begin{abstract}
We explore the nature of the electroluminescence (EL) emission of purely n-doped GaAs/AlGaAs resonant tunneling diodes (RTDs) and the EL evolution with voltage. A singular feature of such a device is unveiled when the electrical output current changes from high to low and the EL on-off ratio is enhanced by 2 orders of magnitude compared to the current on-off ratio. By combining the EL and current properties, we are able to identify two independent impact ionization channels associated with the coherent resonant tunneling current and the incoherent valley current. We also perform the same investigation with an associated series resistance, which induces a bistable electrical output in the system. By simulating a resistance variation for the current-voltage and the EL, we are able to tune the EL on-off ratio by up to 6 orders of magnitude. We further observe that the EL on and off states can be either direct or inverted compared to the tunneling current on and off states. This electroluminescence, combined with the unique RTD properties such as the negative differential resistance (NDR) and high frequency operation, enables the development of high speed functional opto-electronic devices and optical switches.
\end{abstract}

\maketitle

\section{Introduction}
The transport signature of resonant tunneling diodes (RTDs), that is, a peak current density followed by a region of negative differential resistance (NDR)~\cite{Tsu,Tsu1}, enables RTDs to be exploited in a broad range of applications such as THz oscillators, high speed logic gates, light detectors, and thermometers~\cite{Feiginov, DBae, Hartmann, Pfenning1}. Although the heterostructure device layouts of RTDs are rather simple, certain requirements must be achieved in order to produce high quality devices, e.g., low leackage current and large peak to valley current ratio (PVCR)~\cite{Wei}. Apart from being the fastest semiconductor devices used for practical applications~\cite{Oshima2016, Izumi2017}, RTDs are also suitable for opto-electronic uses as they can be light emitters and detectors~\cite{Hartmann, Hartmann2, Romeira2013}. In purely n-doped RTDs, light emission occurs via  electroluminescence (EL) due to either impact ionization processes taking place along the structure~\cite{White, HSheng, Hartmann2} or Zener tunneling, that is, direct interband tunneling from valence to conduction band under high electric fields~\cite{Growden2018, Growden2018_2}. Light emission enables the extraction of reliable information about the diode such as charge build up, carrier dynamics, and system temperature, all of which enrich the whole transport picture~\cite{Yan, Teran, Schwegler}. Being the electroluminescence an intrinsic response of the RTD to electrical excitation, it provides insights on the resonant tunneling process by revealing features of the transport mechanisms of the available charge carriers without the need of an external light source for electron-hole pair creation.

In this paper, we explore the coexistence of two transport channels in a purely n-doped RTD based on GaAs/AlGaAs and assess how they affect the emitted light by contributing independently to the impact ionization processes responsible for hole generation. The experimental observations have been theoretically simulated allowing for the corroboration of the hypothesis raised. We then engineer the EL emission configuration from monostable to bistable, by tuning a resistance in series with the diode. We show that the optical on-off ratio can be larger than the electrical on-off ratio by three orders of magnitude and can be varied within 6 orders by changing the resistance in just one order of magnitude. This architecture can be a tool for developing high quality optical logic devices~\cite{MJiang}.

\section{Sample and Experimental Setup}
\begin{figure}[h!]
\linespread{0.5}
\par
\begin{center}
\includegraphics[scale=0.4]{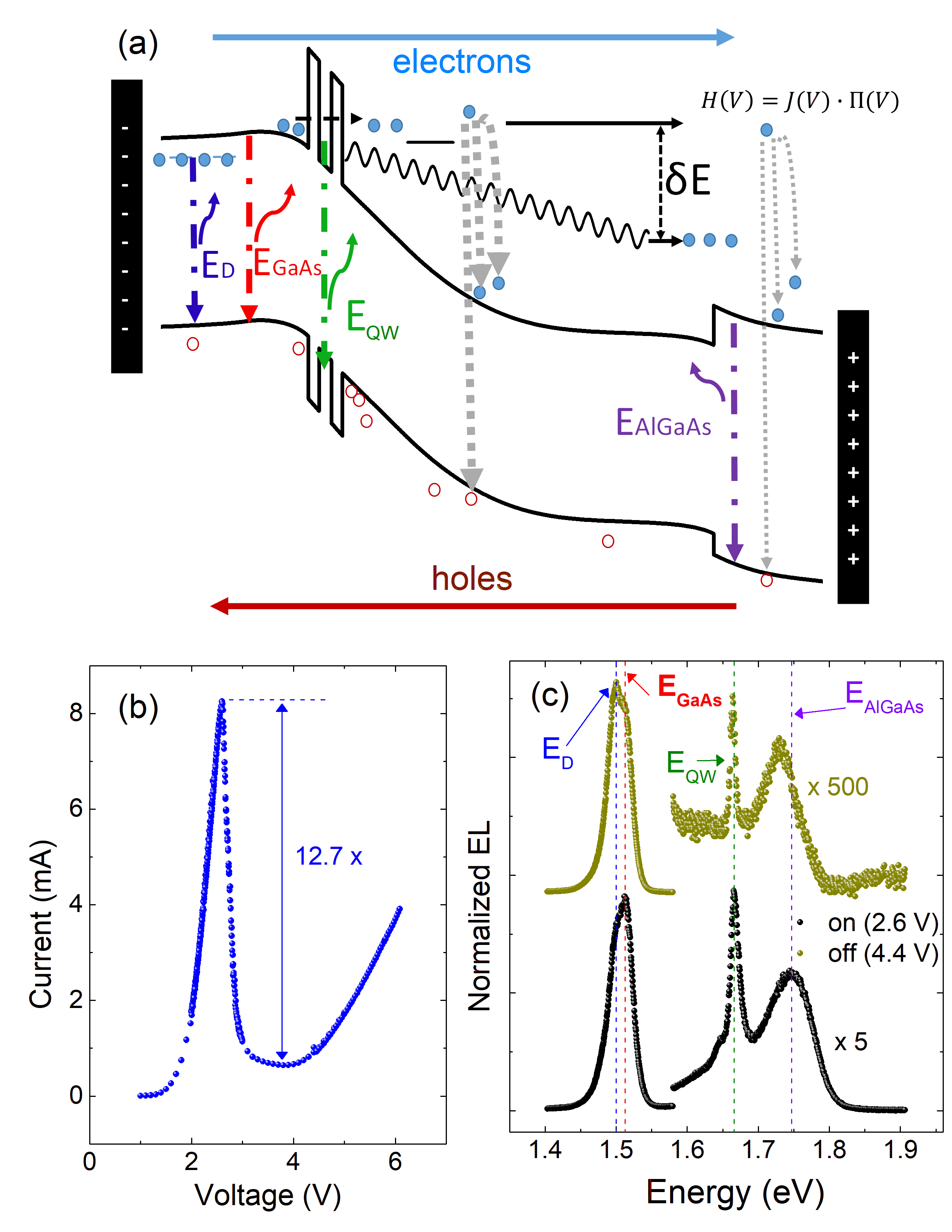}
\end{center}
\par
\vspace{-0.5cm} \caption{(a) Numerical simulation of the band structure under forward bias. The blue (red) horizontal arrow shows the direction of electrons (holes) at this bias regime. Tunneling electrons undergo impact ionization processes either in the GaAs or AlGaAs region (grey arrows, with the ionization probability represented schematically through the arrows' thicknesses) and the generated holes eventually recombine, so the region of each EL emission (vertical colored arrows) is shown schematically. (b) Measured I-V characteristic at T=10 K. (c) Normalized EL spectra obtained at 2.6 V (black) and 4.4 V (dark yellow), respectively before and after the resonance peak.}
\label{bandstructure_el}
\end{figure}

The RTDs were grown by molecular beam epitaxy on top of a Si n-doped ($3\times10^{18}$ cm$^{-3}$) GaAs substrate beginning with a 300 nm n-doped ($3\times10^{18}$ cm$^{-3}$) GaAs buffer layer. The diode consists of an 100 nm n-doped (from $3\times10^{18}$ cm$^{-3}$ to $5\times10^{17}$ cm$^{-3}$) GaAs layer, followed by an undoped region with two 20 nm-thick GaAs layers embedding two 3.5 nm-thick Al$_{0.6}$Ga$_{0.4}$As barriers separated by a 4 nm-thick GaAs quantum well (QW), forming the structure. The RTD is finalized by 100 nm GaAs with a doping concentration of $n=2\times10^{17}$ cm$^{-3}$ followed by 300 nm n-doped (from $2\times10^{17}$ cm$^{-3}$ to $3\times10^{18}$ cm$^{-3}$) Al$_{0.2}$Ga$_{0.8}$As layer, and a 10 nm-thick n-doped ($3\times10^{18}$ cm$^{-3}$) GaAs capping layer. After the growth process, RTD mesas with diameter of 11 $\mu$m were processed by electron beam lithography in combination with dry chemical etching techniques. Fig.~\ref{bandstructure_el}~(a) presents the numerically simulated band profile of the RTD under forward bias, as well as the drift direction of electrons and holes in blue and red arrows, respectively.

Electrical and optical measurements were performed with the sample placed in a helium closed-cycle cryostat (Attocube - Attodry1000). All measurements presented in this study were obtained at a temperature of $T = 10$ K. The setup was used in combination with a confocal microscope (AttoCFMI) to carry out the electroluminescence experiments. The signal was collimated by an objective lens and transmitted along a multimode optical fiber, being dispersed by a 75 cm spectrometer and detected by a silicon charged couple device detector (Andor - Shamrock/Idus).

\section{Results and Discussion}
The I-V characteristic is shown in Fig.~\ref{bandstructure_el}~(b). The peak current is $8.25$ mA at $2.6$ V, and the PVCR is $12.7$. Depending on the spatial origin of the hole generation via impact ionization, the holes drift and recombine with electrons in different regions of the structure, e.g., close to the collector contact~\cite{Hartmann2}, the QW, or the emitter region. Fig.~\ref{bandstructure_el}~(c) shows the normalized EL spectra obtained in two different voltage regimes, before (black circles, bias voltage 2.6 V) and after (dark yellow circles, bias voltage 4.4 V) the resonant current peak. One can observe four distinct emission lines, which are schematically shown in Fig.~\ref{bandstructure_el}~(a) (dot-dashed arrows): $E_{\mathrm{D}}=1.500$ eV - associated with donor-band recombination, $E_{\mathrm{GaAs}}=1.512$ eV is the band to band emission in bulk GaAs, $E_{\mathrm{QW}}=1.664$ eV originates from the quantum well, and $E_{\mathrm{AlGaAs}}=1.746$ eV emerges from the AlGaAs region.

\begin{figure}[h!]
\linespread{0.5}
\par
\begin{center}
\includegraphics[scale=0.4]{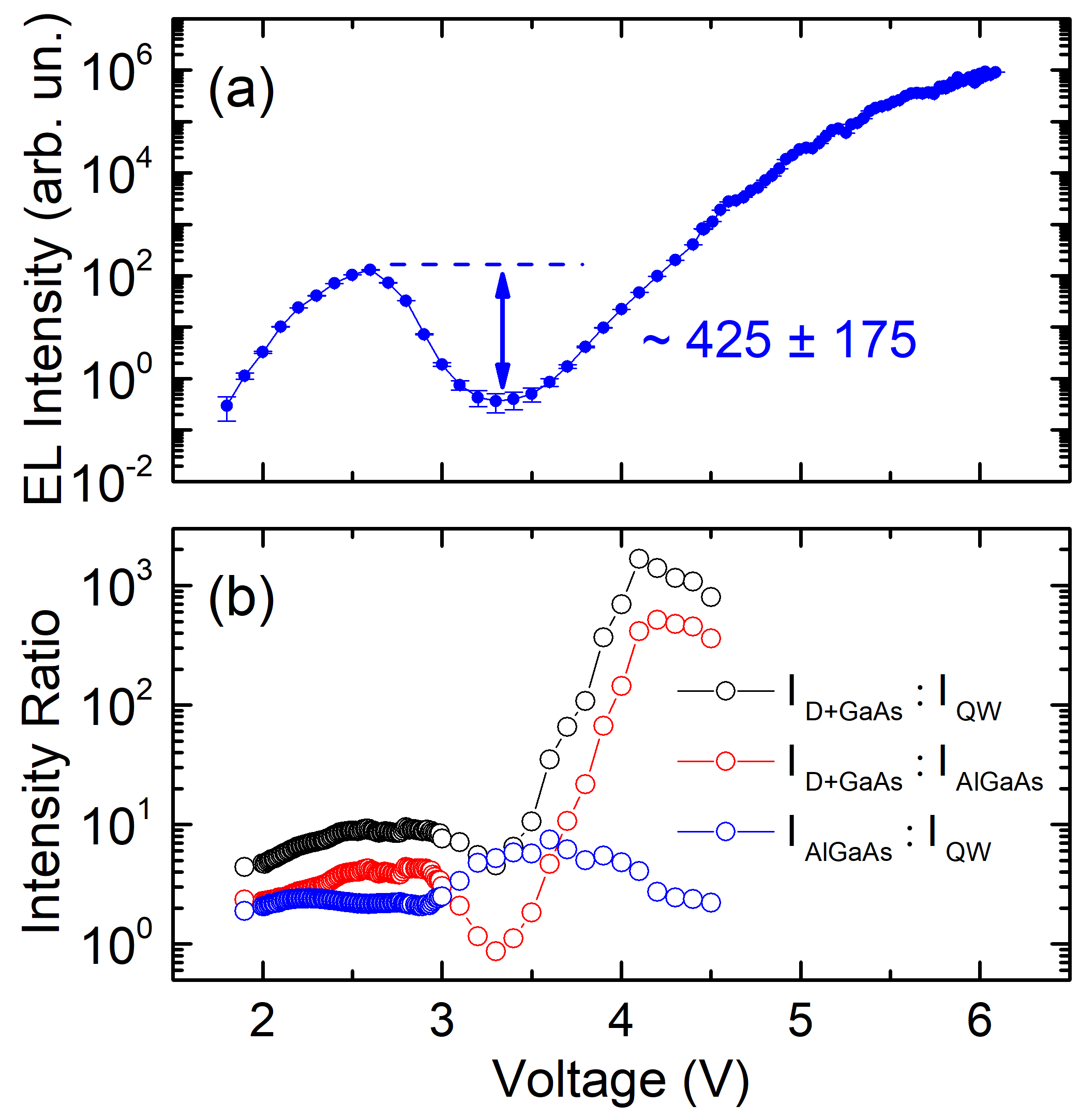}
\end{center}
\par
\vspace{-0.5cm} \caption{(a) EL total intensity as function of voltage. (b) Voltage dependent intensity ratio between GaAs and QW (black), GaAs and AlGaAs (red), and AlGaAs and QW (blue) emissions.}
\label{ixv_el_ratio}
\end{figure}

The QW and AlGaAs emission peaks are weaker compared to $E_{\mathrm{D}}$ and $E_{\mathrm{GaAs}}$ lines, being $\sim5\times$ lower in the on-resonance case (2.6 V), and $\sim500\times$ lower off-resonance (4.4 V). For lower voltages, up to the resonance peak, the charge carrier density in the quantum well increases, leading to a higher recombination rate~\cite{Goodings1994}. Afterwards, in the off-resonance case, the charge carrier density in the quantum well is reduced, increasing the ratio between the bulk GaAs and QW emissions. Additionally, after the current resonance (i.e., at 4.4 V) the impact ionization in the GaAs layer after the double barrier is enhanced due to high electric field in this region, leaving less electrons available to reach the AlGaAs with energies above the impact ionization threshold. The variation with voltage of the intensity integrated over the whole spectrum (EL-V) is presented in Fig.~\ref{ixv_el_ratio}~(a). The EL peak to valley ratio (PVR) is two orders of magnitude higher than the I-V PVCR, with an optical on-off ratio of approximately $425 \pm 175$.

The ratios between the donor emission line plus GaAs ($I_{\mathrm{D+GaAs}}$), AlGaAs ($I_{\mathrm{AlGaAs}}$), and the QW ($I_{\mathrm{QW}}$) intensities are depicted in Fig.~\ref{ixv_el_ratio}~(b). They allow assessing the correlation between the available recombination channels. One can see that the ratio between AlGaAs and QW emission is nearly constant (blue circles) and can be assumed as correlated as the voltage increases. In contrast, when  compared with the GaAs emission, the ratio remains constant before the resonance but suffers a considerable increase after 3.5 V (black and red circles). This is an evidence of a connection between the electrons localized at the resonant state in the QW with those contributing to the AlGaAs emission. The lack of data points beyond 4.5 V is due to limitations of the experimental setup.

\begin{figure}[h]
\linespread{1.0}
\par
\begin{center}
\includegraphics[scale=0.3]{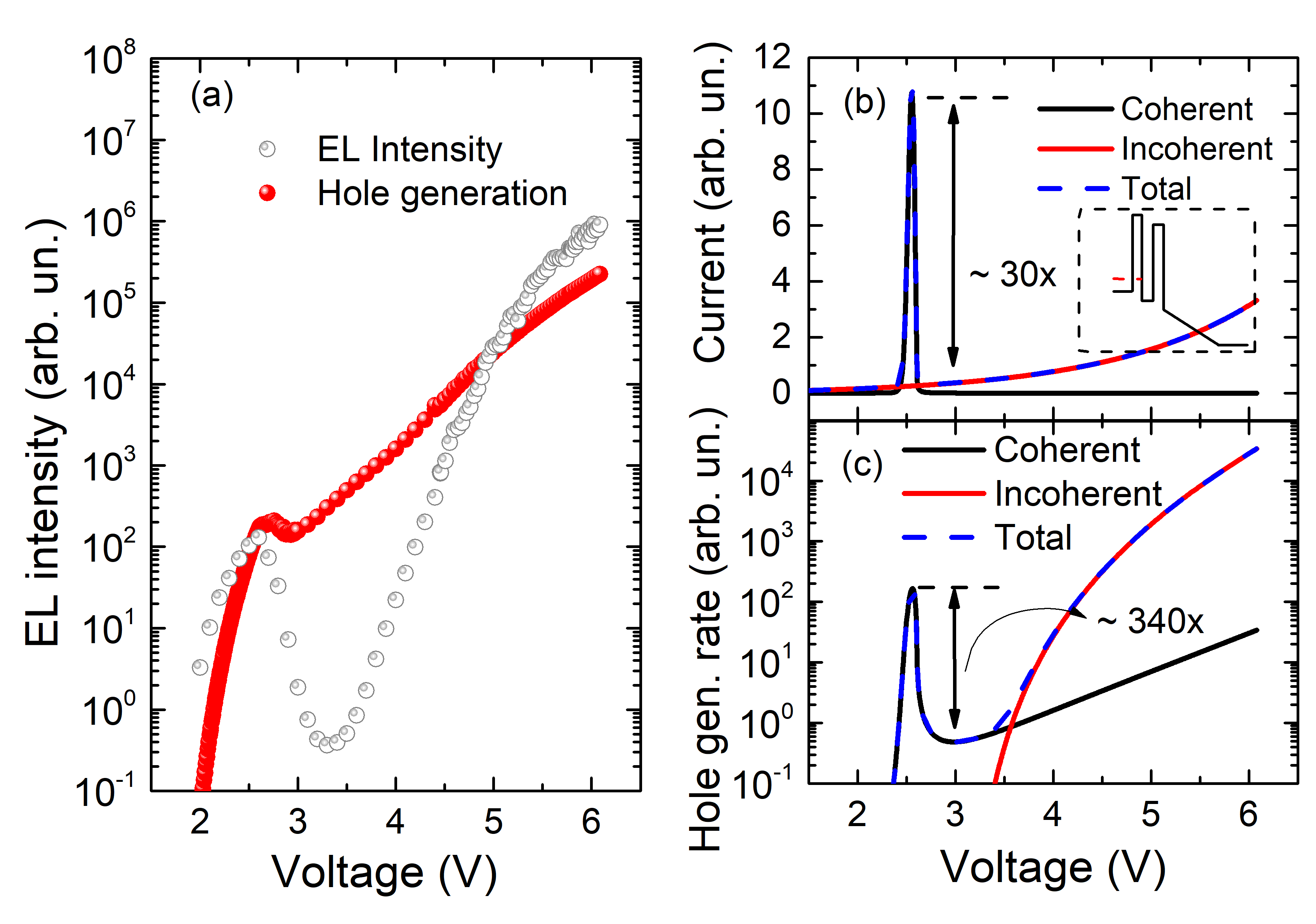}
\end{center}
\par
\vspace{-0.5cm} \caption{(a) EL intensity as function of voltage. Grey circles represent the experimental data, red circles are the product of the current with the impact ionization rate calculated for GaAs. (b) Calculated I-V characteristic, (c) Hole generation rate deduced from the calculated I-V. The inset on (b) represents the bandstructure considered in the model, with a flat band for the barriers and the well.}
\label{ixv_el}
\end{figure}

Another peculiar feature observed in Fig.~\ref{ixv_el_ratio}~(a) is the exponential EL signal growth after the resonance, while the current, in Fig.~\ref{bandstructure_el}~(b), shows a slow increase. As mentioned above, the main process for generating holes in this purely n-doped structure under applied bias is impact ionization, and thus the hole generation rate, $H(V)$, should be proportional to both the current density, $J(V)$, and the impact ionization rate, $\Pi(V)$,
\begin{equation}
H(V)\propto J(V)\Pi(V),
\label{holegen}
\end{equation}
with the ionization rate being described by the generalized Keldysh model~\cite{Redmer}
\begin{equation}
\Pi(V)=C\left(\frac{eV-E_{\mathrm{th}}}{E_{\mathrm{th}}}\right)^{a},
\label{impaction}
\end{equation}
\\
where $C =93.659\times10^{10}$ s$^{-1}$ and $a=4.743$, are derived parameters for GaAs, and $E_{\mathrm{th}}=1.8$ eV is the GaAs impact ionization threshold energy. Our observation matches well with the latter parameter, since the structure starts emitting light from 1.8 V, as one can see in Fig.~\ref{ixv_el_ratio}~(a). By using Eq.~\ref{holegen} with the measured current it is obtained the hole generation rate, equivalent to the total EL intensity, and is presented in Fig.~\ref{ixv_el}~(a) as red circles together with the experimental data (grey circles). The hole generation rate is similar to the EL intensity before the resonance, however it fails from the NDR region and forward.  At first glance, this could be ascribed to electric field corrections, not considered by Eq.~\ref{impaction}. According to Ref.~\onlinecite{Redmer}, when the system is under applied bias, the impact ionization model must be corrected, however the deviations occur at high electric fields and only for lower energies, close to the threshold, $E_{\mathrm{th}}$. Yet, in our case, the same field regime is achieved at higher energies (or voltages), where the models that include the electric field converge to Eq.~\ref{impaction}, so the latter can be considered valid. Thus, a single transport channel contributing to the EL is proven to be inefficient to simulate the RTD emission and it remains to prove that the presence of two impact ionization channels allows explaining this discrepancy.

In RTDs, the current can be divided into a coherent part, $J_{\mathrm{coh}}$, related to the resonant tunneling peak, and an incoherent part. The coherent current flowing through the double barrier structure is simulated using the Tsu-Esaki equation for the density current, given by~\cite{Tsu}
\begin{multline}
J_{\mathrm{coh}}=\dfrac{em^{*}k_{B}T}{2\pi^{2}\hbar^{3}} \int_{0}^{\infty}T(E,V) \\ \times \ln \left[\dfrac{1+\exp\left(\dfrac{E_{f}-E}{k_{B}T}\right)}{1+\exp\left(\dfrac{E_{f}-E-eV}{k_{B}T}\right)}\right]dE,
\label{coh}
\end{multline}
where $e$, $m*$, $k_{B}$, $T$, $\hbar$, $E$, $E_{f}$, $V$ and $T(E,V)$ are, respectively, the electron charge, effective electron mass of the material in the emitter, Boltzmann constant, temperature, reduced Planck constant, electron energy along the growth direction, emitter Fermi level, the voltage necessary to obtain the inner tunneling transmission at the double barrier, and the transmission coefficient. The latter one is obtained by using the transfer matrix method~\cite{Singh} and the band profile under an applied voltage, as shown in the inset of Fig.~\ref{ixv_el}~(b). On the other hand, the incoherent (or excess) part, $J_{\mathrm{inco}}$, comprises other currents such as hot electrons, thermal activation, and sidewall leakage.~\cite{Heiblum,Ternent,Rascol} Thermal activation can be discarded from this model as cryogenic temperatures suppress this process. The sidewall current is dependent on the perimeter of the structure and for higher diameters this contribution becomes less relevant and can also be neglected (which is the case for a diameter of 11 $\mu$m). The excess current is then considered to be composed mostly by electrons with phase coherence and energy losses, and can be accounted by a standard exponential diode term~\cite{Schulmann}, such as
\begin{equation}
J_{\mathrm{inco}}=h\left[\exp\left(\frac{\eta_{1}V}{k_{B}T}\right)-1\right],
\label{inco}
\end{equation}
where $h$ is a fitting parameter and $\eta_{1}$ refers to the efficiency of the excess current.

Following Fig.~\ref{bandstructure_el}~(a), the coherent current consists of carriers tunneling through the double barrier without losing their phase coherence and energy as it is a resonant process, whereas the incoherent current consists of carriers tunneling sequentially into the double barrier, therefore losing their phase coherence and energy through scattering processes such as phonon or lattice scattering. With this in mind, the coherent and incoherent EL channels consist, respectively, of
\begin{equation}
H_{\mathrm{co}}\propto J_{\mathrm{co}}\Pi(V)
\end{equation}
and
\begin{equation}
H_{\mathrm{inco}}\propto J_{\mathrm{inco}}\Pi(V-\delta E),
\end{equation}
\\
with $\delta E$ being the energy loss due to the scattering processes. In a more general picture, a $\delta E$ term could also be added in the expression for the coherent channel, but as the EL emission starts exactly at the threshold energy, $E_{\mathrm{th}}=1.8$, we can assume that no energy loss take place for this channel.

When the coherent channel is off, the incoherent transport, described with Eq.~\ref{inco}, becomes dominant, prevailing in the hole generation for the external bias above 3.5 V. With these assumptions, and using $E_{f} = 5$ meV and T = 10 K, the calculations for the current are presented in Fig.~\ref{ixv_el}~(b), where due to the completely coherent picture, the peak is narrower compared to the experiment. For the hole generation rate, the calculated curve shows good agreement with the EL intensity, as displayed in Fig.~\ref{ixv_el}~(c). Thus, the calculated PVR for the hole generation rate is $\sim 340$, which coincides within the experimental error with the PVR displayed in Fig. \ref{ixv_el_ratio} (a).

\begin{figure}[h!]
\linespread{1.0}
\par
\begin{center}
\includegraphics[scale=0.3]{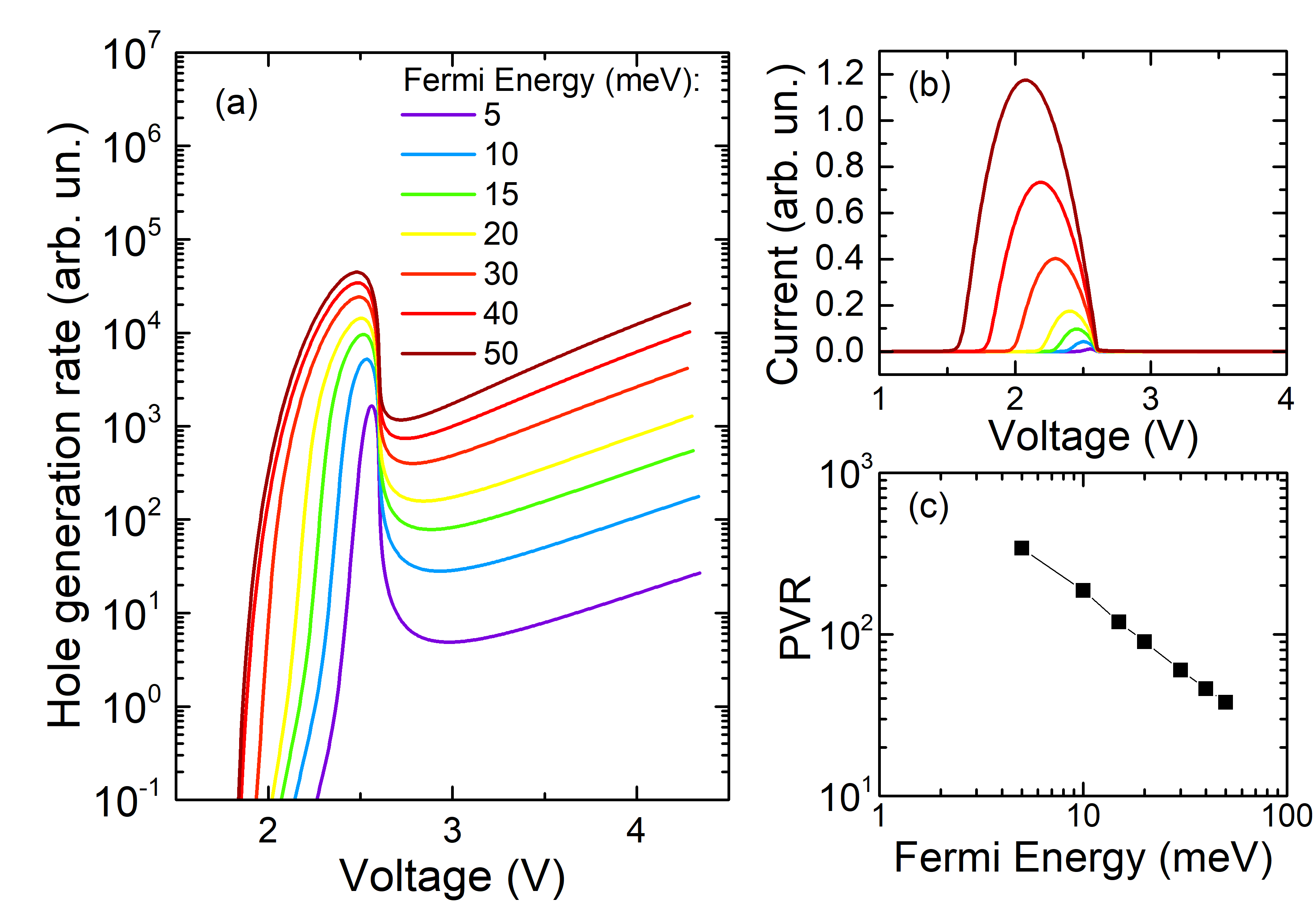}
\end{center}
\par
\vspace{-0.5cm} \caption{(a) Hole generation rate calculated for different Fermi energy values. (b) Simulated current. (c) Hole generation PVR as function of Fermi energy.}
\label{holegen2}
\end{figure}

Furthermore, a value of $\delta E = 1.3$ eV was extracted by adjusting the incoherent channel according to the experiment. Under resonant conditions in RTDs, the contribution of coherent electrons prevails over the incoherent ones~\cite{Rascol}, and, according to Ref.~\onlinecite{Rauch}, their mean free path in GaAs, at an n-doping concentration of $1\times10^{17}$ cm$^{-3}$, is 80 nm. Thus we can assume that these coherent electrons are not likely to undergo any energy relaxation before the impact ionization event. On the other hand, after the resonance, charge carriers are more likely to experience lattice scattering, as the probability of coherent transport drops, which reduces the mean free path and leads to the expected high energy losses for the incoherent electrons. By considering the full picture of coherent and incoherent carriers, the correlation between QW and AlGaAs intensities, shown in Fig~\ref{ixv_el_ratio}~(b), can be understood as follows: only electrons tunneling resonantly through the double barrier can reach the AlGaAs region for generating holes, whereas those subjected to sequential tunneling are not able to go further with energy higher than the impact ionization threshold. The same model was used to assess the effect of Fermi energy variation for a fixed temperature at 10 K, and the results are shown in Fig.~\ref{holegen2}, for the completely coherent case. By increasing $E_{F}$, both the hole generation and the current increase, as depicted in Figs.~\ref{holegen2} (a) and (b), respectively. However such an increase leads to an exponential reduction of the peak to valley ratio, displayed in Fig.~\ref{holegen2}(c).

\begin{figure}[h!]
\linespread{0.5}
\par
\begin{center}
\includegraphics[scale=0.3]{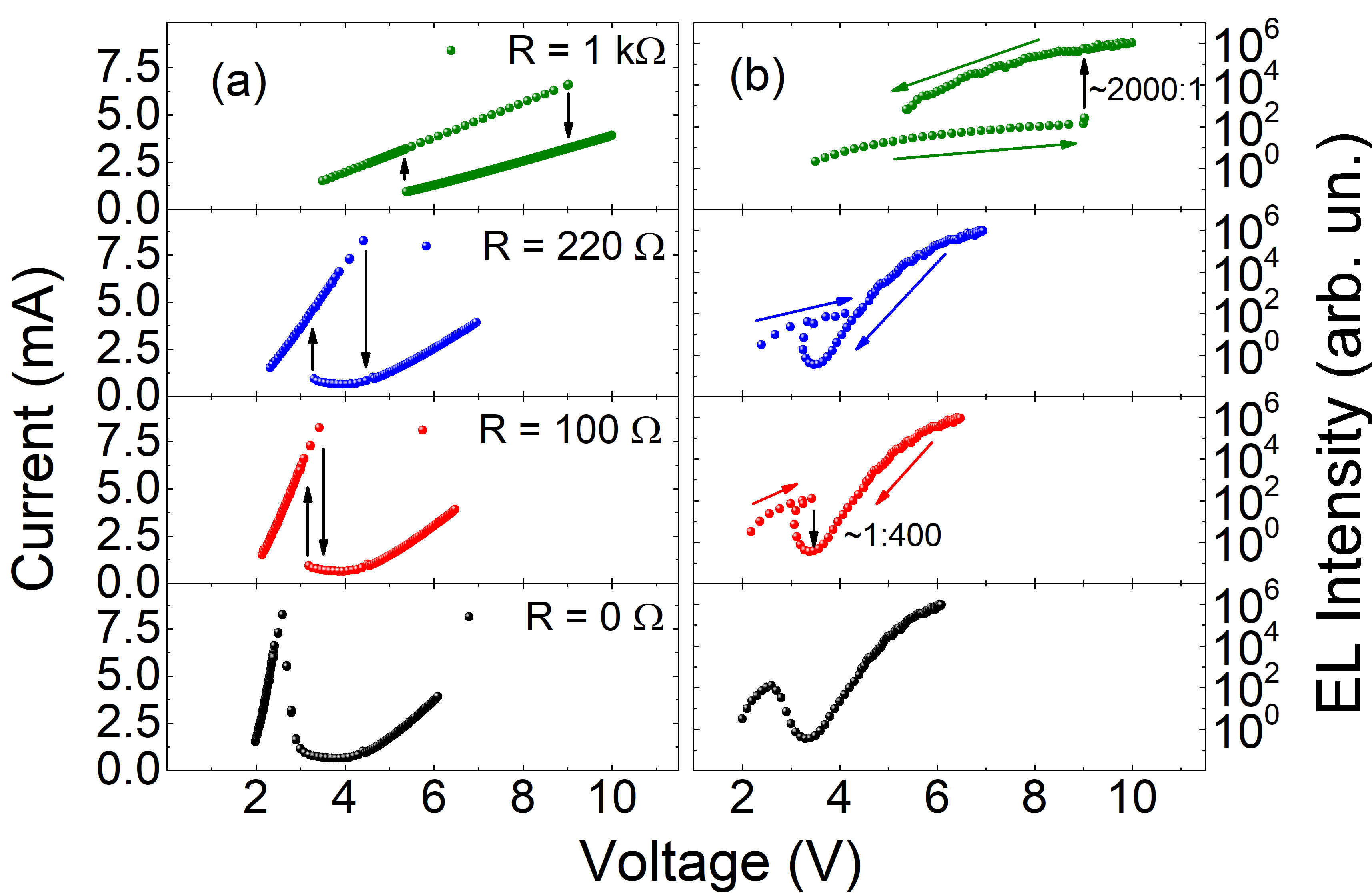}
\end{center}
\par
\vspace{-0.5cm} \caption{(a) I-V and (b) EL-V characteristics. The data presented for 0 $\Omega$ (black circles) and 1 k$\Omega$ (green circles) resistances are experimental while for 100 $\Omega$ and 220 $\Omega$, displayed respectively as red and blue circles, are simulations using Kirchhoff's law.}
\label{sim_res}
\end{figure}

Once the nature of the modulation of the light emission is understood, we can proceed and discuss how to engineer the optical response.
A desirable feature in optical logic devices is the thorough control over the emission enhancement an decrease, which is useful to develop optical trigger-related functionalities.~\cite{MJiang,Hu} One may think, for instance, in changing between on/off or off/on optical states by simply varying the associated resistance. Figs. \ref{sim_res}~(a) and (b) show, respectively, the I-V characteristics and EL-V, for four different resistances in series. The data presented for 0 $\Omega$ (black circles) and 1 k$\Omega$ (green circles) were obtained experimentally. The effects of intermediary values of the resistance can be simulated by using Kirchhoff's law and the experimental $I(V)$ function without resistance association as starting point. In this case, the total bias is defined by $V_{ext}=V+I(V) \cdot R$ and the results for $R=100$ $\Omega$ and $R=220$ $\Omega$ are displayed as red and blue circles, respectively, in Figs. \ref{sim_res}~(a) and (b).

\begin{figure}
\linespread{0.5}
\par
\begin{center}
\includegraphics[scale=0.3]{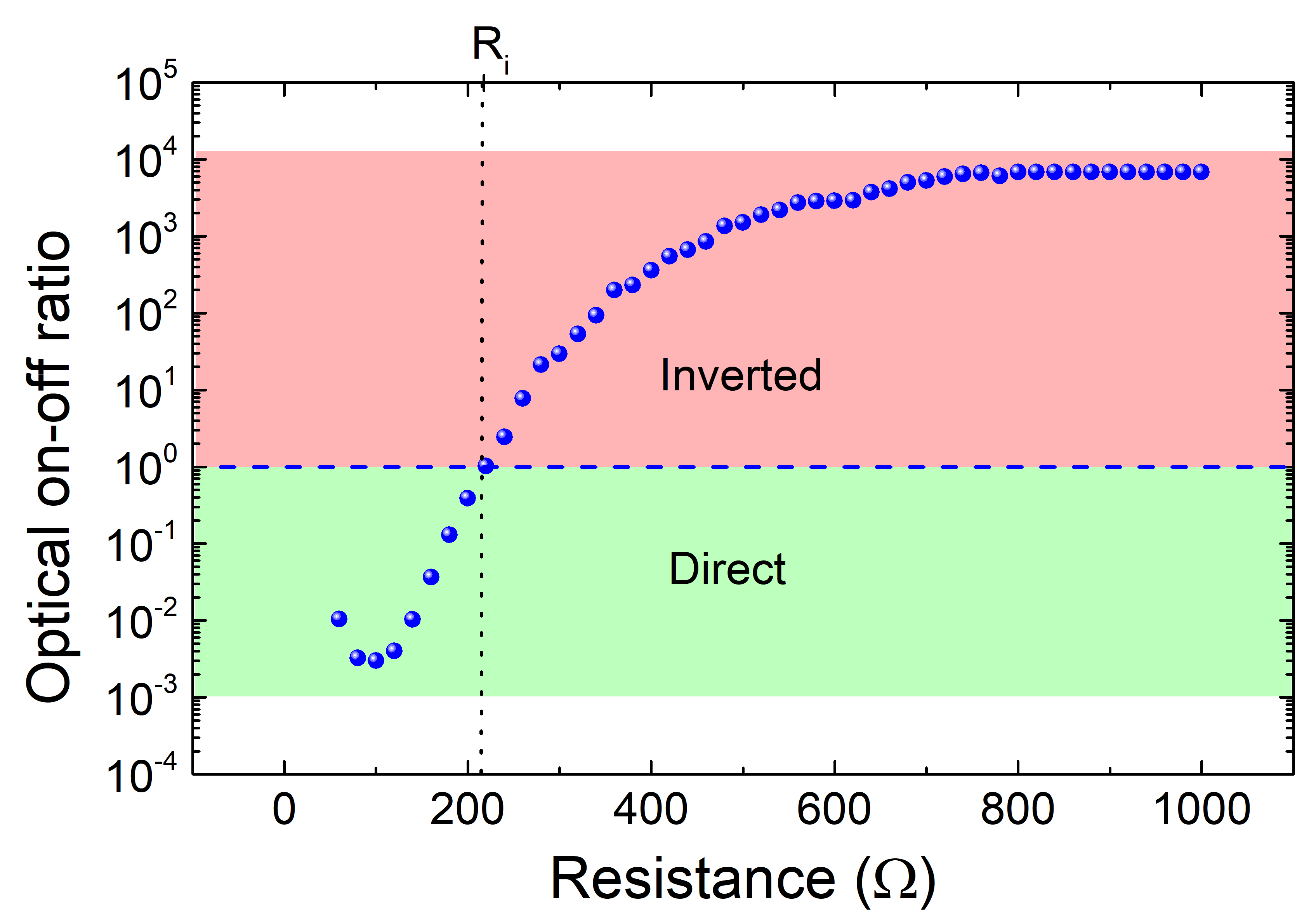}
\end{center}
\par
\vspace{-0.5cm} \caption{Simulated optical on-off ratio versus series resistance. The green and red regions represent direct and inverted switch compared to the current-voltage transition, respectively. The resistance at which the inversion occurs is $R_{i}=220$ $\Omega$.}
\label{on-off-ratio}
\end{figure}

The direction of the voltage sweep has been indicated with arrows in Fig.~\ref{sim_res}. Note that for $R=100$ $\Omega$, the bistable course of the luminescence changes the light state in the ratio of $\sim 1:400$ right after the resonance and the EL bistability follows the same current path. However, for $R=1000$ $\Omega$, the EL bistability is inverted with respect to the current and the measured on/off ratio has grown up to $2000:1$.
Fig.~\ref{on-off-ratio} displays the calculated optical on-off ratio for a set of simulated resistances from 0 up to 1 k$\Omega$ by using the Kirchhoff's law, where the conditions for direct and inverted ratios can be assessed. The ratio is increased with higher resistances until an inversion point at $R_{i}=220$ $\Omega$, and then it goes up to $\sim10000:1$. This means that we can, in principle, modulate up to 7 orders the light emission enhancement by changing just the resistance. The experimental on-off ratio with 1 k$\Omega$ resistance association is $\sim2000:1$. Possible reasons for this discrepancy, where the simulated ratios appear overestimated, are: a non-linear internal resistance or a sensitivity increase due to the high resistance, caused by noise or fluctuations. Nevertheless, our results clearly show the modulation of the emitted light by 6 orders of magnitude.

\section{Conclusions}
The correlation between the coherent and incoherent transport channels and the hole generation rate has been assessed in a standard purely n-doped GaAs/AlGaAs-based RTD. Each of the coherent and incoherent current channels has been proven to contribute with independent impact ionization processes. Electrons being transported through the incoherent channel undergo a relatively high energy loss due to lattice scattering. A series resistance has been successfully used as a simple yet effective tool to tune both the current and electroluminescence emission versus inducing bistable states. By simulating a resistance variation over the I-V characteristics we have shown that it is possible to tune the intensity of the emission by up to 6 orders of magnitude, and the on-off switch can be either direct or inverted compared to the transport on-off states. These EL properties, such as high emission and tunability, could be useful for trigger-related and optical logic devices.

\section{Acknowledgment}
The authors gratefully acknowledge the financial support of the following agencies: FAPESP (grants $\#$ 2013/18719-1, 2014/07375-2, 2014/19142-1, 2014/02112-3), BAYLAT-FAPESP (grant $\#$  2016/50080-9 ), CNPq (grants $\#$ 141793/2015-5, 471191/2013-2), CAPES (grant $\#$88881.133567/2016-01) and the German Ministry of Education and Research (BMBF) within the national project HIRT (FKZ13XP5003B).

\bibliographystyle{apsrev}

\bibliography{bibliography}

\end{document}